\documentclass[
    nobibnotes, aps, preprint,
    superscriptaddress,
    obeyspaces, spaces,
]{revtex4-2}

\makeatletter
\def\switch@array{}
\makeatother

\usepackage{multirow}
\usepackage{xcolor}
\usepackage{tabularx}
\usepackage{ragged2e}

\usepackage{lmodern}                
\usepackage[T1]{fontenc}            
\usepackage[utf8]{inputenc}         

\usepackage{amsmath}
\usepackage{graphicx}               
\usepackage{longtable}
\usepackage{makecell}               
\usepackage{booktabs}               
\usepackage{dcolumn}                

\usepackage{longtable}
\usepackage{booktabs}
\usepackage{tabularx}
\usepackage{ragged2e}
\usepackage{array}
\newcolumntype{P}[1]{p{#1}}  

\usepackage[
    colorlinks,
    citecolor=blue,
    urlcolor=blue,
    filecolor=blue,
    linkcolor=black
]{hyperref}

\usepackage{tcolorbox}
\tcbuselibrary{breakable}

\usepackage{soul}                   
\usepackage{color}                  
\usepackage{xspace}                 

\begin{document}

\newcommand{\cseaff}{
School of Computational Science and Engineering, Georgia Institute of Technology, 771 Ferst Drive NW, Atlanta, GA 30332, USA
}

\newcommand{\mseaff}{
School of Materials Science and Engineering, Georgia Institute of Technology, 771 Ferst Drive NW, Atlanta, GA 30332, USA
}

\newcommand{\matmerizeaff}{
Matmerize, Inc., Atlanta, GA 30308, USA
}

\title[]{
    Retrieval Augmented Generation of Literature-derived Polymer Knowledge: The Example of a Biodegradable Polymer Expert System
}

\author{Sonakshi Gupta}
    \affiliation{
        School of Computational Science and Engineering,
        Georgia Institute of Technology, 771 Ferst Drive NW, Atlanta 30332,
        GA, USA.
    }
\author{Akhlak Mahmood}
    \affiliation{
        Matmerize, Inc., Atlanta 30308,  
        GA, USA.
    }
\author{Wei Xiong}
    \affiliation{\mseaff} 
\author{Rampi Ramprasad}
    \email{rampi.ramprasad@mse.gatech.edu}
    \affiliation{\mseaff}
    \email{rampi.ramprasad@mse.gatech.edu.}


\begin{abstract}

Polymer literature contains a large and growing body of experimental knowledge, yet much of it is buried in unstructured text and reported with inconsistent terminology, making it difficult to retrieve, compare, or reason over systematically. Existing tools often extract narrow, study-specific facts, without providing the integrated context needed to answer scientific questions that span multiple studies. Retrieval-augmented generation (RAG) offers a promising way to overcome this limitation by combining large language models (LLMs) with external retrieval, but its effectiveness depends strongly on how domain knowledge is represented. In this work, we develop and adapt two retrieval pipelines for a materials science corpus: a dense semantic vector-based approach (VectorRAG) and a graph-based approach (GraphRAG). Using over 1,000 polyhydroxyalkanoate (PHA) papers, we construct context-preserving paragraph embeddings and a canonicalized structured knowledge graph that supports entity disambiguation and multi-hop reasoning. To evaluate these pipelines, we conduct extensive benchmarking using retrieval metrics, comparisons with general state-of-the-art systems such as GPT and Gemini, and qualitative validation by a domain chemist. The results show that GraphRAG achieves higher precision and interpretability, while VectorRAG provides broader recall, underscoring complementary trade-offs between the two approaches. Expert validation further confirms that the tailored pipelines, particularly GraphRAG, produce well-grounded, citation-reliable responses with high domain relevance. By grounding every statement in explicit source evidence, these systems offer researchers a dependable way to navigate the literature, compare findings across studies, and uncover patterns that are difficult to extract manually. More broadly, this work demonstrates a practical path for developing scholarly assistants in materials science using curated corpora and retrieval design, reducing dependence on large proprietary models and enabling accessible, trustworthy literature analysis at scale.
\end{abstract}


\maketitle
\raggedbottom

\section*{Main}
Data has long been recognized as a cornerstone of materials informatics \cite{agrawal2016perspective, ramprasad2017machine, himanen2019data}, and this has become increasingly evident as recent advances in the field continue to rely on data-driven techniques. From algorithm design to autonomous discovery workflows, recent breakthroughs have been enabled not only by improvements in machine learning, but also by the availability of large, high-quality materials datasets. For many polymer systems, however, such datasets remain limited or unavailable. Instead, relevant information is primarily reported through a vast, heterogeneous, and largely unstructured body of literature that continues to expand at an unprecedented pace, with decades of scientific knowledge dispersed across text, figures, and tables \cite{martin2023emerging}. As a result, much of this information is difficult to retrieve, integrate across studies, or use effectively for scientific reasoning.

Previous efforts have sought to tap into this reservoir \cite{gupta2024data, shetty2023, cheung2024polyie, kalhor2024functional, bai2024schema}, producing structured data that can support downstream tasks such as property prediction \cite{gupta2025benchmarking, stergiou2023enhancing}, synthesis planning \cite{chen2021data, kim2017materials}, and materials design \cite{savit2025polybart, sahu2025encoder}. While these databases provide valuable static snapshots of extracted knowledge, they do not fully address how polymer scientists interact with the literature in practice. Scientific inquiry often requires dynamically retrieving evidence across studies, reconciling inconsistent terminology, and generating results reported under varying experimental contexts. Consequently, even seemingly straightforward questions such as “What gas permeability values have been reported for polyhydroxyalkanoate films under different processing conditions?” or “What products form during PHB hydrolysis under acidic versus alkaline conditions?” cannot typically be answered by consulting a single paper or database record. Instead, addressing such queries requires identifying relevant experimental contexts and integrating fragmented findings reported across multiple sources, underscoring the need for systems that support literature-grounded reasoning rather than isolated fact extraction.

An effective polymer literature scholar must therefore go beyond extraction to (i) retrieve relevant evidence across papers, (ii) synthesize information while preserving experimental context, (iii) appropriately defer when the literature does not provide sufficient evidence, and (iv) explicitly ground conclusions in verifiable source material so that domain experts can assess reliability and trustworthiness.

\begin{figure*}
    \centering
    \includegraphics[width=6.4in]{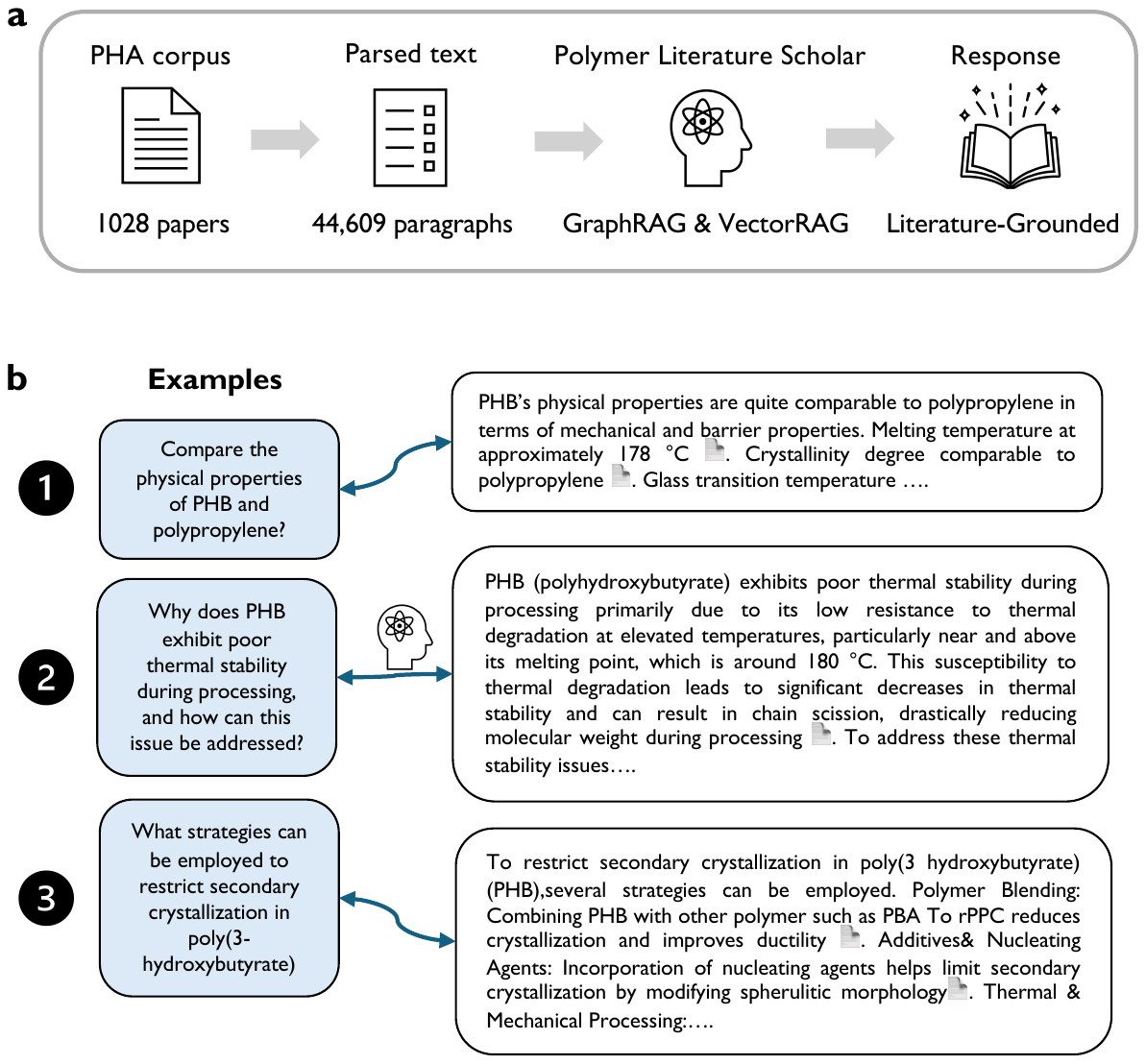}
    \caption{(a) Overview of the retrieval-augmented generation (RAG) framework for literature-grounded question answering on PHAs. The curated PHA corpus, comprising 1,028 articles and 44,609 parsed paragraphs, serves as the shared knowledge base for both VectorRAG and GraphRAG retrieval pipelines. Retrieved contextual evidence from each pipeline is provided to a LLM to generate literature-grounded responses. 
    (b) Example scientific queries and corresponding responses generated by the Polymer Literature Scholar. Responses are synthesized by aggregating evidence retrieved from multiple relevant studies within the PHA corpus, as indicated by the document symbol.}
    \label{fig:fig1}
\end{figure*}

Large language models (LLMs) offer powerful capabilities for synthesizing dispersed scientific information, owing to their ability to leverage knowledge learned from vast text corpora. However, a well-documented limitation of LLMs is hallucination, wherein they produce fluent and confident but factually incorrect statements \cite{augenstein2024factuality, alansari2025largelanguagemodelshallucination}. Recent work \cite{kalai2025languagemodelshallucinate} shows that hallucinations arise because LLMs are optimized to complete text rather than to validate information, and standard training objectives reward confident answers even when the underlying evidence is weak. This behavior becomes particularly concerning in scientific settings, where incorrect statements can be difficult to detect without explicit grounding. To mitigate hallucinations and improve reliability, researchers have explored a range of strategies, including prompt engineering, refining the model's output during generation, self-feedback mechanisms, and retrieval-augmented generation (RAG) approaches \cite{tonmoy2024comprehensivesurveyhallucinationmitigation, shim2025multistagepromptrefinementmitigating, madaan2023selfrefineiterativerefinementselffeedback, Ayala_2024}. In RAG systems, models retrieve relevant information from trusted external sources and condition their responses on this evidence, reducing unsupported claims while improving transparency and traceability; making RAG particularly well suited for scientific domains.

RAG methods have evolved rapidly in recent years, driven by advances that integrate retrieval more tightly with generation and enable models to incorporate external knowledge more effectively \cite{fan2024surveyragmeetingllms}. These developments have produced two prominent retrieval paradigms: vector-based dense retrieval and graph-based reasoning \cite{ahmad2025benchmarking}. Vector-based RAG (VectorRAG) represents information as dense embeddings in a high-dimensional space and retrieves semantically similar passages through similarity search, with extensive work focused on optimizing chunking, embedding models, indexing, and scoring mechanisms \cite{10603291}. Graph-based RAG (GraphRAG), in contrast, organizes knowledge as a network of interconnected entities and relations, enabling retrieval through graph traversal or subgraph extraction. This structured representation supports multi-hop reasoning, improves retrieval precision in domains with complex relational dependencies, and produces evidence trails that are more interpretable \cite{peng2024graphretrievalaugmentedgenerationsurvey}. Together, these complementary retrieval paradigms provide the foundations needed to support reliable and grounded LLM reasoning in materials science, where both accurate evidence retrieval and interpretable relational reasoning are essential. 

Adapting these retrieval strategies to materials science introduces unique challenges. Scientific text in the field is highly heterogeneous, marked by complex jargon, nested property–composition–processing relationships, and inconsistencies in nomenclature. For example, the same polymer may be described as “PLA,” “poly(lactic acid),” or “polylactide,” making it harder to bring together and compare results reported across different studies. These challenges are particularly pronounced for polyhydroxyalkanoates (PHAs), a diverse and rapidly expanding family of biodegradable polymers of significant interest for sustainable materials design \cite{olonisakin2025recent}. As such, PHAs provide a compelling test case for evaluating complementary retrieval strategies such as VectorRAG and GraphRAG.

In this work, building on the framework illustrated in Fig. 1, we introduce the Polymer Literature Scholar, a retrieval-augmented system designed to support literature-grounded reasoning for polymer science.  We develop and systematically benchmark VectorRAG and GraphRAG pipelines on a curated corpus of over 1,000 full-text PHA articles, tailoring each approach to the domain. VectorRAG is adapted through domain-aware chunking and embedding strategies, and GraphRAG through robust entity extraction, normalization, and canonicalization pipelines that support scalable graph reasoning. To systematically assess retrieval quality and factual grounding, we curated a benchmark of 113 domain-expert evaluation questions focused on PHAs and conducted aqualitative expert analysis spanning general scientific queries, paper-specific questions, and multi-paper reasoning tasks. The responses generated by five representative systems: GraphRAG with GPT-4o-mini, GraphRAG with Llama-3.1-70B, VectorRAG with GPT-4o-mini, OpenAI’s built-in RAG with ChatGPT-5 (Web Search), and Google’s built-in RAG with Gemini were evaluated based on content accuracy, citation completeness, and depth of scientific reasoning. 

To support practical use and expert assessment of the system, we also developed an interactive interface exposing both the VectorRAG and GraphRAG retrieval pipelines and their underlying evidence. This interface allows domain experts to query the system, inspect the retrieved paragraphs or knowledge graph tuples, verify evidence provenance, and assess the credibility and completeness of the generated responses.

A key contribution of this work is demonstrating how state-of-the-art retrieval methods can be adapted to build a reliable, domain-focused Polymer Literature Scholar using only a carefully curated corpus and modest computational resources, rather than the scale or expense typically associated with general-purpose foundation models. By presenting a transparent and reproducible retrieval-augmented framework, we show that reliable polymer literature reasoning does not require large proprietary models. Even with smaller, open-weight models, the system remains effective, explicitly grounds its answers in the literature, and appropriately responds with “I do not know” when the available evidence is insufficient. Using PHAs as an illustrative example, we show how vector-based and graph-based retrieval can be aligned with the structure of polymer science literature. VectorRAG provides broad semantic coverage and captures paragraph level context effectively, while GraphRAG supports more structured, relationship aware reasoning across linked entities. Together, these approaches offer a grounded and interpretable basis for reasoning over complex and inconsistently reported polymer data, making literature-driven polymer research both practical and credible.

\section*{RESULTS AND DISCUSSION}
\subsection*{Overview of the Retrieval-Augmented Reasoning Pipelines}
Fig.~\ref{fig:fig1} provides an overview of the Polymer Literature Scholar, a retrieval-augmented framework and interactive application designed to support scientific reasoning over the experimental literature on PHAs. Rather than retrieving isolated facts, the system enables the synthesis of coherent, literature-grounded explanations that reflect how polymer properties, processing behavior, and structure–property relationships are discussed and interpreted across multiple studies.

As shown in Fig. 1a, the framework is built on a curated corpus of 1,028 full-text journal articles focused on PHA chemistry, processing, and properties. These articles are parsed into 44,609 paragraph-level text segments, which serve as the fundamental units of scientific evidence. This paragraph-level representation preserves experimental context that is central to how materials scientists reason about polymer behavior. To enable effective access to this literature, the Polymer Literature Scholar supports two complementary retrieval strategies: VectorRAG and GraphRAG.

In the VectorRAG pipeline, paragraph-level text is embedded into a continuous semantic space, allowing retrieval of experimentally relevant passages even when similar concepts are described using different terminology across papers. In contrast, the GraphRAG pipeline organizes extracted entities such as polymer systems, processing steps, and measured properties into a structured representation that explicitly captures relationships reported in the literature. This graph-based view enables reasoning across interconnected statements, allowing evidence to be integrated from multiple studies that collectively describe a material behavior or trend.

During inference, a user interacts with the Polymer Literature Scholar through a developed application interface by submitting a query. Relevant evidence is retrieved using either of the two retrieval pathway, and the resulting paragraphs or subgraphs are provided as contextual input to a large language model. The model then synthesizes a response grounded in experimentally reported observations drawn directly from the PHA corpus, enabling users to obtain literature-supported explanations.

\begin{figure*}
    \centering
    \includegraphics[width=6.48in]{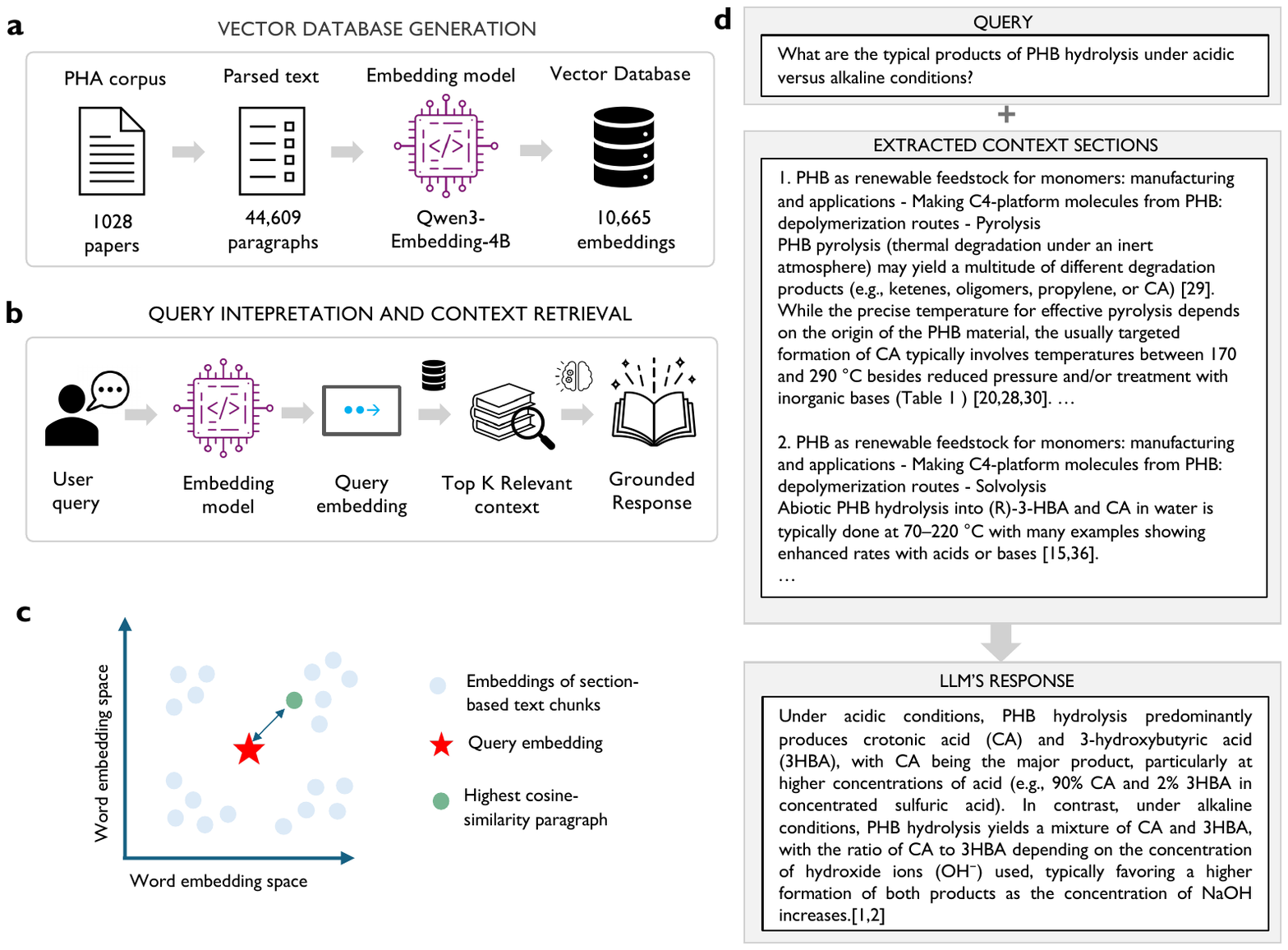}
    \caption{\textbf{VectorRAG workflow for literature-grounded question answering on PHAs.}
    (\textbf{a}) Backend processing of the corpus, where full-text articles are parsed, contextually grouped into condensed text chunks, and embedded into a dense semantic space for similarity-based retrieval. 
    (\textbf{b}) Retrieval-augmented inference, in which user queries are encoded in the same latent space and matched to the most semantically aligned text segments from the corpus. 
    (\textbf{c}) Conceptual representation of the embedding space illustrating how relevant paragraphs are identified through cosine similarity between the query vector and its nearest neighbors. 
    (\textbf{d}) Example of a representative query showing retrieved literature passages and the corresponding grounded, citation-linked response generated by the language model.}
    \label{fig:fig2}
\end{figure*}

Figure 1b illustrates representative example queries and responses generated through the interface. Rather than citing a single source, each response reflects aggregated evidence from multiple relevant studies, as indicated by the document symbol in the figure. This mode of aggregation mirrors how materials scientists typically engage with the literature, by integrating observations across papers to develop a consensus understanding of material behavior, while maintaining transparency about the evidentiary basis of each response. As a result, the framework allows scientists to query the literature in a way that aligns with experimental practice, enabling literature-grounded insights into polymer properties, processing considerations, and design choices that are difficult to obtain using conventional keyword search or database-driven tools.

\subsection*{VectorRAG: Dense Semantic Retrieval Pipeline}
The VectorRAG framework represents the literature corpus in a high-dimensional semantic space capable of capturing contextual similarity across studies. As illustrated in Fig.~\ref{fig:fig2}, the workflow integrates three main components: knowledge base construction, semantic similarity–based retrieval, and grounded answer generation.

In the knowledge base construction stage (Fig.~\ref{fig:fig2}a), the 44,609 parsed paragraphs are grouped into context-preserving text chunks by concatenating adjacent paragraphs from the same section or subsection. This approach ensures that text segments describing related concepts such as synthesis steps and resulting properties remain contiguous, and results in 10,665 context-preserving text chunks. These chunks are embedded using the \textit{Qwen3-Embedding-4B} model to produce 3,584-dimensional dense vectors, which are then stored in a relational vector database for efficient similarity-based retrieval.

During query processing (Fig.~\ref{fig:fig2}b–c), a user query is embedded into the same vector space and compared with all corpus embeddings using cosine similarity. The number of retrieved segments ($k$) was varied between 2 and 8, with optimal grounding and response coherence observed at $k = 8$. The retrieved  chunks are concatenated with the user query and provided to an LLM, which generates a literature-grounded response supported by explicit citations (Fig.~\ref{fig:fig2}d).

This dense semantic retrieval approach effectively captures implicit conceptual relationships and broader narrative context, making it particularly effective for descriptive or exploratory questions such as “What factors influence the crystallinity of PHBV copolymers?” However, at the same time, the unstructured nature of retrieved text can introduce redundancy and higher token utilization during inference. VectorRAG therefore provides a strong semantic retrieval baseline that complements the structured, relational reasoning capabilities of GraphRAG.

\begin{figure*}
    \centering
    \includegraphics[width=6.48in]{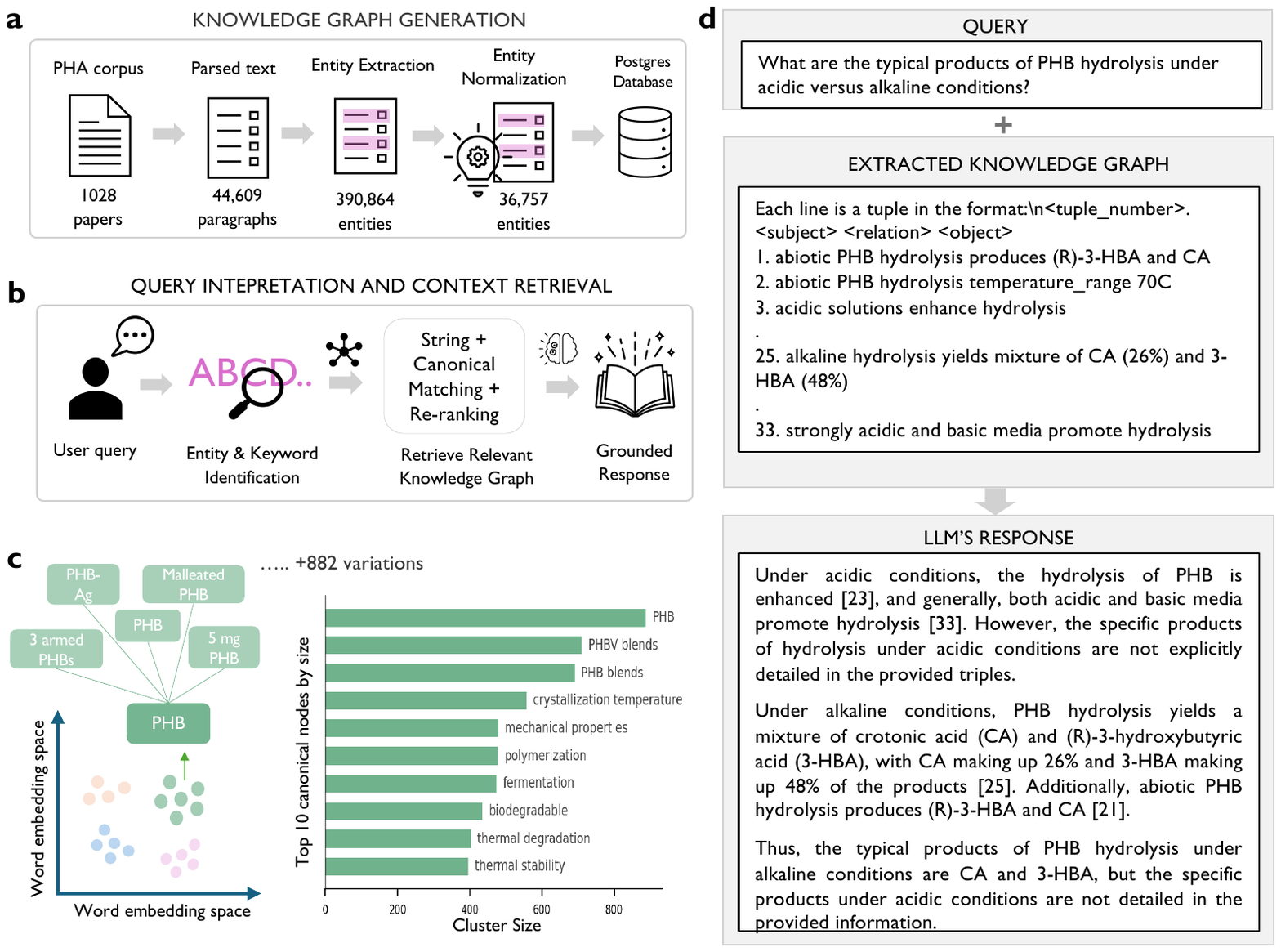}
    \caption{\textbf{GraphRAG workflow for knowledge graph–based question answering on PHAs.} 
    (\textbf{a}) Backend processing of the corpus, where entities and relationships are extracted from literature, normalized, and stored as relational tuples in a relational database. 
    (\textbf{b}) Retrieval and reasoning stage showing how user queries are decomposed into entity–relation pairs, matched to canonical entities, and re-ranked through a path-based scoring strategy to identify the most relevant subgraphs for grounded response generation. 
    (\textbf{c}) Example of entity canonicalization, where multiple related mentions (e.g., PHB–Ag, maleated PHB, 3-armed PHB) are merged into a unified canonical node representing PHB. The accompanying bar plot highlights the top canonical entities ranked by cluster size, demonstrating how normalization improves graph connectivity and recall. 
    (\textbf{d}) Representative example showing a user query, retrieved knowledge graph tuples, and the corresponding grounded response synthesized by the language model with supporting citations.}
    \label{fig:fig3}
\end{figure*}

\subsection*{GraphRAG: Knowledge Graph–Based Retrieval Pipeline}
The GraphRAG framework restructures the parsed corpus into an explicit knowledge graph (KG), enabling interpretable, multi-hop reasoning across diverse relationships within scientific text. As illustrated in Fig.~\ref{fig:fig3}, the workflow comprises three main components: knowledge graph construction, query interpretation and retrieval, and grounded response generation.

In the backend construction phase (Fig.~\ref{fig:fig3}a), the 44,609 parsed paragraphs are processed through an entity extraction and normalization pipeline, as described in the Methods section. Several LLMs were evaluated, including GPT-4o, GPT-4o-mini, GPT-4.1-mini, Llama-3.1-70B, and Llama-3.3-70B, to assess how model choice influences tuple quality and downstream retrieval. Among these, GPT-4o-mini and Llama-3.1-70B produced the most reliable tuples and were therefore selected for the final pipeline. The extraction prompts and all variations tested during this stage are provided in the Supporting Information.

Consequently, GPT-4o-mini extracted 390,864 relational tuples and Llama-3.1-70B extracted 311,475 tuples. These tuples capture scientific relationships such as \texttt{[PHB-has property-Tg]} or \texttt{[PHBV-synthesized with-hexanoate]}. Because polymers and related entites often appear under multiple textual variants, an entity normalization step was introduced to merge equivalent mentions and increase graph connectivity. This was applied to merge equivalent mentions and render the graph more interconnected. This step yielded 36,757 canonical entities for GPT-4o-mini and 37,289 for Llama-3.1-70B, improving consistency and retrieval quality across the corpus. All resulting tuples were stored in a relational database, creating a structured, queryable representation of scientific knowledge derived from the PHA literature.

During query processing (Fig.~\ref{fig:fig3}b), each user question is decomposed into its key entities and relations using an entity and keyword parser. GraphRAG retrieves relevant subgraphs by combining canonical entity matching with string and semantic similarity, as described in the Methods section. The retrieved tuples are then re-ranked through the path-reranking strategy and capped at a maximum subgraph size of 300 nodes. These subgraphs often contain multiple paths linking related entities, thereby enabling multi-hop reasoning across studies. For example, a query such as “How does 3HV content affect the melting temperature of PHBV copolymers?” triggers traversal from \textit{PHBV → 3HV → Tm}, retrieving all relevant experimental contexts associated with this relationship.

As shown in Fig.~\ref{fig:fig3}c, the graph consolidates textual variations into canonical nodes, where contextually related entities (e.g., “3-armed PHB,” “malleated PHB,” and “PHB–Ag”) are unified under a single canonical entity, PHB. The right-hand panel of Fig.~\ref{fig:fig3}c highlights the top canonical nodes ranked by cluster size, corresponding to the most frequently discussed entities, such as PHB, PHBV blends, crystallization temperature, and degradation behavior.

Finally, the retrieved subgraph context is concatenated with the user query and passed to the LLM for citation-grounded response generation (Fig.~\ref{fig:fig3}d). By representing evidence as explicit relational tuples, GraphRAG provides a transparent reasoning chain that enhances explainability and traceability. Compared with VectorRAG, this structured retrieval paradigm emphasizes relational precision and reasoning depth, making it particularly effective for mechanistic or comparative questions that require multi-hop inference across studies.

\subsection*{Benchmarking and Evaluation}
\subsubsection*{Controlled Evaluation: Benchmarking Retrieval Fidelity}

As an initial test, a subset of 21 PHA-related DOIs was used to develop and optimize both retrieval pipelines. This controlled setup ensured that the RAG framework could effectively retrieve relevant information from a focused dataset where every paper was directly related to the query topics. A total of 113 questions were constructed manually after reviewing the full texts of these 21 DOIs to evaluate retrieval fidelity using the Recall and Accuracy metrics described in the Methods section.

The 21 full-text articles were parsed into 943 paragraphs, which were embedded and stored in the vector database for the VectorRAG pipeline. In parallel, entity extraction and normalization produced 10,814 raw tuples, consolidated into 5,904 canonical entity nodes using GPT-4o-mini, and 8,744 raw tuples consolidated into 4,973 canonical nodes using Llama-3.1-70B, for the GraphRAG pipeline.  These datasets formed the benchmark for testing retrieval precision and grounding quality prior to scaling up to the complete corpus.

\begin{table}[htbp]
\centering
\caption{\textbf{Controlled Evaluation}: Reported metrics include recall, accuracy, and average per-query response time and cost for GraphRAG and VectorRAG on a subset of 21 PHA DOIs, using optimized context limits of 300 tuples and 8 paragraphs, respectively.}
\renewcommand{\arraystretch}{1.3} 
\setlength{\tabcolsep}{8pt} 
\begin{tabular}{lccccc}
\toprule
\textbf{Models} &
\makecell{\textbf{Context}\\\textbf{limit}} &
\textbf{Recall} &
\makecell{\textbf{Accuracy}} &
\makecell{\textbf{Avg. response}\\\textbf{time (s)}} &
\makecell{\textbf{Avg. total}\\\textbf{cost (\$)}} \\
\midrule

\multicolumn{6}{l}{GraphRAG} \\
GPT-4o-mini & 300 tuples& 0.982 & 0.991 & 7.18 & 0.00097 \\
Llama-3.1-70B & 300 tuples& 0.991 & 0.991 & 10.23 & 0.00000 \\

\midrule
\multicolumn{6}{l}{VectorRAG} \\
GPT-4o-mini & 8 paragraphs& 0.973 & 1.000 & 16.43 & 0.00186 \\
Llama-3.1-70B & 8 paragraphs& 0.973 & 1.000 & 41.75 & 0.00000 \\

\bottomrule
\end{tabular}
\label{tab:test1_results}
\end{table}

As shown in Table~\ref{tab:test1_results}, both retrieval pipelines performed exceptionally well on the focused 21-DOI test set, achieving near-perfect recall across all models. The high recall values indicate that, when the corpus is narrowly defined and thematically consistent, both dense (VectorRAG) and structured (GraphRAG) retrieval frameworks can successfully retrieve all relevant information. GraphRAG achieved faster retrieval and lower computational cost owing to its compact, relational representation, whereas VectorRAG exhibited higher latency primarily due to processing substantially longer text inputs, often approaching the 8,192-token context limit of the model. These results demonstrate that both pipelines can effectively ground model responses under controlled conditions and establish a reliable baseline for subsequent large-scale evaluations.

\subsubsection*{Scaled Evaluation: Retrieval Performance at Corpus Scale}
To assess scalability and generalization beyond the controlled benchmark, both retrieval pipelines were evaluated on the complete PHA corpus comprising of 1,028 full-text journal articles. Performance was measured using recall and accuracy, complemented by average per-query response time and total inference cost to capture efficiency, as summarized in Table~\ref{tab:test2_results}.

When scaled to the full corpus, both pipelines exhibited a decline in recall, reflecting the increased difficulty of retrieving the precise context as the search space expanded. The reduction was more pronounced for VectorRAG, with a recall of 0.717, suggesting that dense semantic retrieval becomes less discriminative when the number of embeddings increases and contextual overlap grows. In contrast, GraphRAG maintained substantially higher recall, 0.938 with GPT-4o-mini, benefiting from its structured, entity-linked retrieval that confines searches to semantically consistent regions of the knowledge graph. Despite these differences, the overall answer accuracy remained consistently high for both frameworks, approximately 0.96-0.97, indicating that even when the expected paragraph was not retrieved, the pipeline often produced correct, literature-grounded responses. This observation highlights that retrieval precision alone does not fully dictate downstream answer quality. Instead, GraphRAG’s relational representation provides greater robustness and interpretability at scale, underscoring the advantages of structured retrieval for large, heterogeneous scientific corpora.
\begin{table}[htbp]
\centering
\caption{\textbf{Evaluation at Scale}: Reported metrics include recall, accuracy, and average per-query response time and cost for GraphRAG and VectorRAG on 1028 PHA DOIs, using optimized context limits of 300 tuples and 8 paragraphs, respectively.}
\renewcommand{\arraystretch}{1.3} 
\setlength{\tabcolsep}{6pt} 
\begin{tabular}{lccccc}
\toprule
\textbf{Models} &
\makecell{\textbf{Context}\\\textbf{limit}} &
\textbf{Recall} &
\makecell{\textbf{Accuracy}} &
\makecell{\textbf{Avg. response}\\\textbf{time (s)}} &
\makecell{\textbf{Avg. total}\\\textbf{cost (\$)}} \\
\midrule

\multicolumn{6}{l}{GraphRAG} \\
GPT-4o-mini & 300 tuples & 0.938 & 0.973 & 34.14 & 0.00070 \\
Llama-3.1-70B & 300 tuples & 0.903 & 0.964 & 24.18 & 0.0 \\

\midrule
\multicolumn{6}{l}{VectorRAG} \\
GPT-4o-mini & 8 paragraphs& 0.717 & 0.960 & 17.69 &  0.00184 \\
Llama-3.1-70B & 8 paragraphs& 0.717 & 0.960 & 40.35 &  0.0 \\

\bottomrule
\end{tabular}
\label{tab:test2_results}
\end{table}


\subsection*{Qualitative and Expert Evaluation}

\subsubsection*{Detailed Analysis of Representative Queries}

To qualitatively assess retrieval and reasoning behavior, a set of representative polymer science questions were analyzed, as shown in Table~\ref{tab:detailed_analysis}. The table presents side-by-side responses from the GraphRAG and VectorRAG pipelines for questions spanning three distinct reasoning categories: mechanistic understanding, process–structure correlation, and property comparison.

In Query~1, both pipelines produced literature-consistent and chemically accurate answers. However, VectorRAG provided a more detailed mechanistic explanation by elaborating on reaction conditions specifically under acidic and alkaline environments demonstrating its advantage when relevant information is captured at the paragraph level. In contrast, GraphRAG generated a concise but less comprehensive response; while correct, it omitted additional contextual details that would otherwise have contributed to a richer set of graph tuples.

In Query~2, GraphRAG delivered a more structured and hierarchical explanation, identifying five major categories of challenges in producing low-density PLA foams and citing relevant literature. This highlights its strength in multi-hop reasoning and information aggregation, linking dispersed evidence across multiple sources into a coherent narrative. The VectorRAG response, though factually accurate, was comparatively brief and lacked descriptive depth.

In Query~3, VectorRAG outperformed GraphRAG by providing quantitative thermal and mechanical property data for P(3HB) and P(4HB). GraphRAG’s response, while limited in scope due to knowledge sparsity within the retrieved subgraph, remained fully factual and free from hallucination, reflecting a key advantage of graph-grounded reasoning in maintaining reliability even when coverage is incomplete. This again emphasizes VectorRAG’s ability to leverage paragraph-level semantic understanding to generate more detailed, context-rich answers.

Overall, the complementary performance of the two pipelines underscores their distinct retrieval paradigms: 
GraphRAG enhances interpretability through structured reasoning and contextual aggregation while maintaining factual precision, whereas VectorRAG excels in producing fluent, paragraph-level explanations when detailed textual information is present in the corpus.

\begingroup
\linespread{0.9}\selectfont

\begin{longtable}{|>{\raggedright\arraybackslash}p{.15\textwidth}|>{\raggedright\arraybackslash}p{.40\textwidth}|>{\raggedright\arraybackslash}p{.40\textwidth}|}
\caption{Qualitative comparison of GraphRAG and VectorRAG responses to representative polymer science questions, illustrating differences in retrieval behavior, reasoning depth, and grounding style between the two pipelines.}

\label{tab:detailed_analysis}\\
\hline
\textbf{Question} & \textbf{GraphRAG Response} & \textbf{VectorRAG Response} \\
\hline
\footnotesize{1. What are the main monomeric products of abiotic PHB hydrolysis and under what conditions are they typically formed?} &
\footnotesize{The main monomeric products of abiotic PHB hydrolysis are (R)-3-HBA and CA [4]. This process typically occurs under conditions of 70–220~°C [5].} &
\footnotesize{The main monomeric products of abiotic PHB hydrolysis are D-(-)-3-hydroxybutyric acid (3HBA) and crotonic acid (CA). These products are typically formed under acidic or alkaline conditions, where the ester linkages in the polymer chain undergo hydrolysis. Specifically, at neutral pH (7.4), 3HB is a major intermediate product of PHB degradation [1]. Under acidic conditions (0.1–4.0 M H), 3HBA and CA are formed, with CA being the dominant monomeric product in highly concentrated acid solutions (80–98 wt\% H2SO4) [2]. Conversely, alkaline conditions (0.1–4.0 M OH-) result in higher amounts of both 3HB and CA.} \\
\hline

\footnotesize{2. What challenges are associated with producing low-density PLA foams, and how can these be addressed?} &
\footnotesize{Producing low-density PLA foams presents several challenges, primarily related to the material's intrinsic properties and processing conditions. 
\par\vspace{6pt}
1. \textbf{Melt Strength and Elasticity:} PLA exhibits low melt elasticity and low melt strength, which complicates the production of low-density foams. These properties can hinder the ability to achieve the desired foam structure during processing [211][212].

2. \textbf{Crystallinity and Foamability:} High crystallinity in PLA can limit the expansion of the produced foam, reducing its foamability [200]. Additionally, slow crystallization kinetics contribute to the challenges in achieving uniform cell structures and desirable densities [266]. } 

3. \textbf{Cell Structure Uniformity:} Achieving uniform cell structures in large-scale settings is particularly challenging for low-density PLA foams. This uniformity is critical for maintaining consistent mechanical properties throughout the foam [213]. & 

{\footnotesize{Producing low-density PLA foams is challenged by factors such as PLA's poor melt strength and low melt elasticity, which lead to inadequate foaming behavior without modifications. 
\par\vspace{6pt}
Additionally, the crystallization behavior of PLA can hinder foaming, particularly at higher crystallization rates, which can prevent effective cell formation and growth. The $CO_{2}$ induced crystallization at high pressures can further complicate the foaming process by increasing crystalline regions that reduce foamability.
\par\vspace{6pt}
These challenges can be addressed by blending PLA with polyhydroxybutyrate (PHBV), which improves the rheological properties and enhances foamability. PHBV acts as a nucleating agent for PLA, potentially leading to better cell structure and reduced density in the foams.}} \\

\hline
& \footnotesize{
\par\vspace{4pt}
4. \textbf{Surface-to-Volume Ratios:} Producing foams with high surface-to-volume ratios is another challenge, as it impacts the foam's overall performance and applications [214]. 

5. \textbf{Processing Variables:} The interaction between material properties and processing conditions is crucial in determining foam characteristics. Variations in pressure, temperature, and the type of foaming agents used can significantly affect the final foam structure [220][250].
\par\vspace{6pt}
To address these challenges, several strategies could be employed:
-- \textbf{Additives and Blending:} Incorporating additives or blending PLA with other polymers, such as PHBV, can enhance the foaming properties and mechanical strength of the resulting foams [18][130]. 

\textbf{Optimized Processing Techniques:} Utilizing advanced processing techniques, such as film/foam coextrusion, may enhance the production of low-density PLA foams by allowing better control over processing parameters [21][264].
} & 

\footnotesize{
\par\vspace{4pt}
Adjusting the composition of the blend and optimizing processing conditions, such as foaming temperature and pressure, can also help achieve the desired foam properties while minimizing the adverse effects of crystallization and melt strength limitations [1][2][4][8].} \\
\hline

\footnotesize{\par\vspace{4pt}
3. How do the physical and thermal properties of P(3HB) and P(4HB) differ and what are the implications of these differences for their potential applications?} &

\footnotesize{\par\vspace{4pt}
The knowledge graph does not provide specific details on the physical and thermal properties of P(3HB) and P(4HB) directly. However, it does mention the analysis of PHB, which is related to P(3HB) and has implications for thermal behavior and processing. 

1. PHB (a homopolymer of 3HB) is analyzed by thermal analysis to understand its properties [4]. 

2. The difference between melting and thermal degradation temperatures affects the processability of PHB production [20]. 

3. Blending with other polymers can improve mechanical properties, flexibility/ductility, and processability of PHAs [14]. } & 

\footnotesize{\par\vspace{4pt}
The physical and thermal properties of P(3HB) and P(4HB) differ significantly, which has implications for their potential applications. P(3HB) is a highly crystalline, water-insoluble polymer with a high Young's modulus (3.5~GPa) and low tensile strength (43~MPa). It is brittle, with an elongation at break of only 5\%, making it less suitable for applications requiring flexibility [3]. Its melting temperature is around 180~°C, and it possesses a glass transition temperature of approximately 4~°C [5]. These characteristics limit its use primarily to applications where rigidity and stability are needed. 
}\\
\hline
 & \footnotesize{\par\vspace{4pt}
While these points provide some context about the thermal properties of PHB, they do not directly compare P(3HB) and P(4HB) or provide their specific physical properties and implications for applications.

Therefore, there is insufficient information in the knowledge graph to answer the question regarding the differences between P(3HB) and P(4HB) and their implications comprehensively.} & 

\footnotesize{\par\vspace{4pt}
In contrast, P(4HB) is a strong, malleable thermoplastic with a tensile strength comparable to polyethylene and an elongation at break of 100\% [4]. This high elasticity makes P(4HB) suitable for applications requiring flexibility and resilience. Its thermal properties include a lower melting point of about 164~°C and a glass transition temperature of 3~°C [6], which allows for easier processing and adaptability in various applications.
\par\vspace{4pt}
The differences in mechanical properties imply that while P(3HB) may be used in rigid applications, such as packaging or structural components, P(4HB) is better suited for applications requiring elasticity and flexibility, such as in biomedical devices or flexible films. Additionally, the ability to create copolymers, such as P(3HB-co-4HB), allows for tailored properties that can merge the strengths of both polymers for diverse applications [8].
} \\

\hline
\end{longtable}
\endgroup

\subsubsection*{Expert Validation}
\begin{figure*}
    \centering
    \includegraphics[width=4.68in]{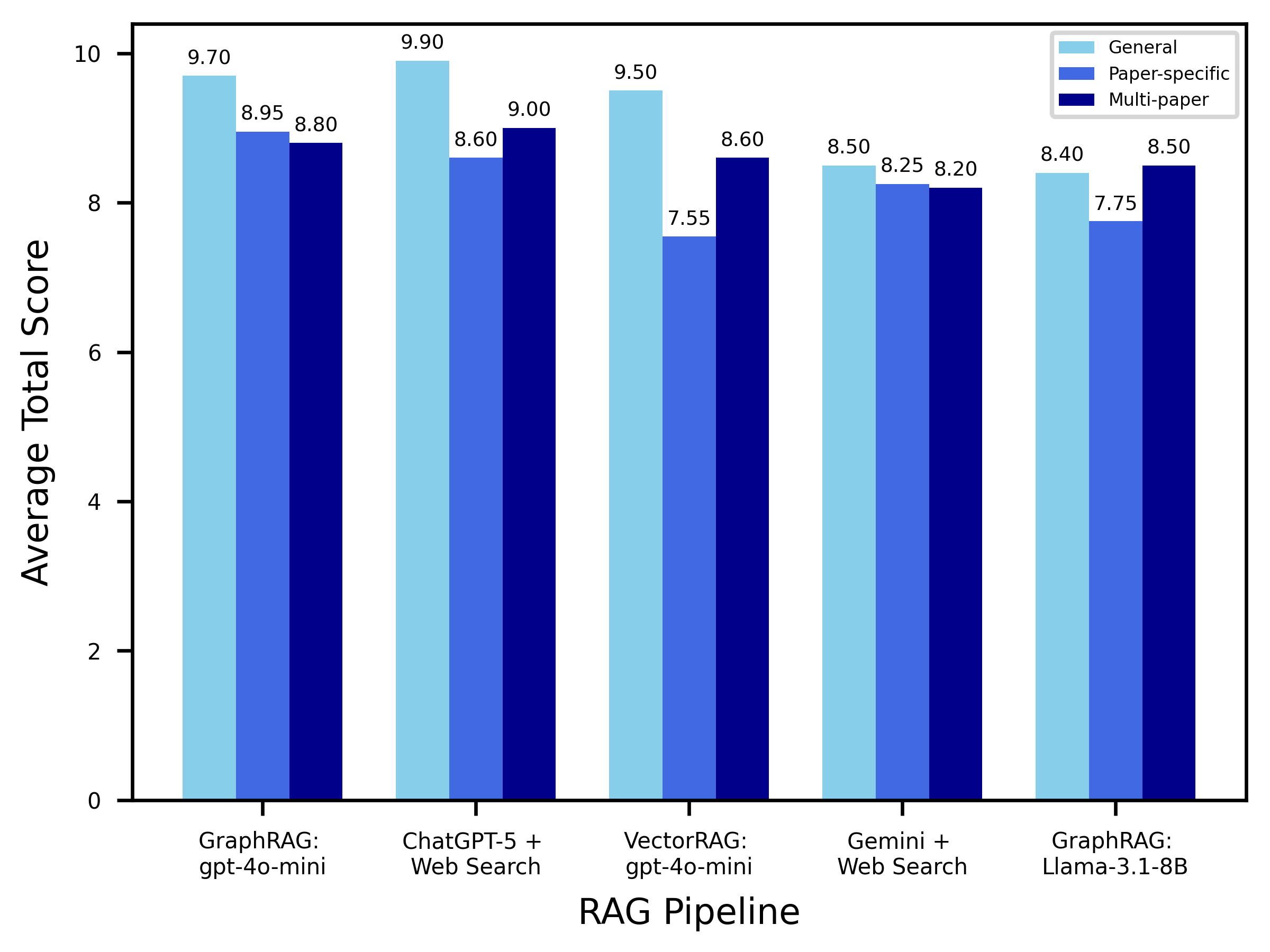}
    \caption{Domain-expert evaluation of five RAG pipelines across General, Paper-specific, and Multi-paper questions. Scores reflect how well each system balanced factual grounding, contextual coverage, and citation reliability. GraphRAG (GPT-4o-mini) and ChatGPT-5 with Web Search achieved the highest overall performance, with other pipelines showing moderate but consistent results.}
    \label{fig:rag_comparison}

\end{figure*}

To better understand how retrieval architecture influences reasoning quality, a domain chemist independently evaluated the performance of five retrieval–generation pipelines using the developed interactive RAG interface. The analysis focused on how each system balances factual grounding, contextual coverage, and citation reliability across different question types. 

The comparison included two variants of the GraphRAG framework built on GPT-4o-mini and Llama-3.1-70B, VectorRAG, ChatGPT-5 with Web Search, which integrates OpenAI’s in-built retrieval-augmented setup combining web evidence with citation tracing, and Gemini with Web Search, Google’s corresponding RAG implementation that retrieves supporting content from online sources. Each model was tested on a three-tier question set, General, Paper-specific, and Multi-paper, designed to reflect the increasing complexity of scientific reasoning, from conceptual clarification to interpretation of individual studies and finally to synthesis across multiple publications. Details of the question set are provided in the Supporting Information. For each answer, the expert assigned a score from one to ten, with five points allocated to content quality and factual correctness and five points allocated to citation quality and relevance.

As shown in Figure~\ref{fig:rag_comparison}, GraphRAG powered by GPT-4o-mini and ChatGPT-5 with Web Search achieved the highest overall scores, averaging between nine and ten across all question categories. Both demonstrated strong factual grounding, coherent reasoning, and reliable citation behavior. The GraphRAG variant built on Llama-3.1-70B scored slightly lower but maintained consistent factual precision, illustrating how language-model capacity influences the quality and completeness of the knowledge-graph tuples and subsequently the downstream answer quality. VectorRAG and Gemini with Web Search produced moderate yet stable performance across all categories as well. A detailed comparison of representative responses is provided in the Supporting Information.

Building on these results, the expert evaluation also revealed clear and consistent differences in how each system reasons over the literature. GraphRAG, particularly when paired with GPT-4o-mini, produced the most precise and well supported answers. Its responses were tightly anchored to the extracted relational evidence and showed very low rates of hallucination, a property that is essential for scientific use. ChatGPT-5, supported by web retrieval, generated more expansive narrative answers, reflecting its broader access to information, but its citations were sometimes less specific than those produced by the in house systems. VectorRAG performed well for questions that depend on paragraph level context, often supplying richer explanations when the needed information was concentrated in a few passages. Gemini offered wider coverage but less reliable referencing due to the variable quality of retrieved web sources. The GraphRAG with Llama-3.1-70B, although more literal, still produced accurate and well grounded answers, demonstrating that the framework remains effective even with smaller open weight models.

An additional and important observation was that the in-house pipelines routinely responded with “I do not know” when the extracted knowledge did not contain sufficient information to answer the question. This behavior stood in contrast to the commercial systems, which were more likely to provide speculative or incomplete answers. The tendency to abstain rather than hallucinate is particularly valuable for scientific applications, where incorrect claims can mislead downstream interpretation.

Taken together, these results show that retrieval pipelines built on a carefully curated knowledge base and domain aware processing can match, and in several respects surpass, the performance of web driven proprietary systems. At the same time, they offer greater degree of transparency, scientific rigor, and substantially lower computational cost. Unlike commercial interfaces that often return links to entire articles, leaving users to manually locate the supporting evidence, the pipelines developed here provide precise evidence trails at the level of individual paragraphs or knowledge graph tuples. This reduces the effort required for verification, strengthens trust in the generated responses, and offers researchers a more reliable foundation for interpreting and analyzing the literature across the polymer domain.

\section*{Conclusion}
Recent advances in LLMs have shown remarkable promise for scientific knowledge discovery, yet their usefulness ultimately depends on the quality and structure of the information they draw from. This work demonstrates that a reliable literature scholar for polymer materials does not require large proprietary systems or opaque retrieval engines. Instead, carefully curated corpora combined with domain-aware retrieval pipelines can provide a foundation that is both trustworthy and tailored to the scientific questions materials researchers need to ask.

By focusing on PHAs as a representative case study, we show how state-of-the-art retrieval methods can be adapted to reflect the structure and reporting practices of polymers literature. Dense semantic retrieval through VectorRAG and structured relational retrieval through GraphRAG both benefit from careful corpus construction, systematic transformation of text, and robust entity normalization, which together create the basis required for LLMs to reason reliably over heterogeneous scientific content. These design choices make the resulting answers traceable, interpretable, and scientifically credible.

The results also reveal a division of strengths across the two retrieval pipelines. VectorRAG delivers broad semantic recall and rich paragraph-level context, making it effective for descriptive or exploratory questions. GraphRAG excels at multi-hop, relational reasoning with compact, interpretable evidence trails. At corpus scale, GraphRAG maintains higher recall and lower latency while VectorRAG often provides more detailed narrative answers when the relevant information is localized in text. Together, they highlight the value of combining complementary retrieval paradigms in a single workflow.

Across controlled benchmarks and expert review, both pipelines achieve high accuracy, with GraphRAG in particular matching the performance of commercial web‑connected RAG systems on domain questions while exhibiting more conservative, evidence-driven behavior and more consistent citations. In particular, the in-house pipelines more frequently defer with “I do not know” when the available literature does not support a definitive answer, reducing the risk of speculative or unsupported claims. At the same time, the study shows that conventional retrieval metrics such as recall cannot fully capture the scientific usefulness of a RAG system. Lower recall does not necessarily indicate weak grounding or limited relevance. Human expertise remains essential for determining whether retrieved context is truly relevant, whether reasoning chains are scientifically sound, and whether citations are complete and trustworthy. 

Taken together, this work provides a practical and reproducible pathway for developing cost-effective and transparent AI scholars tailored to the needs of the materials science community. By grounding model reasoning in curated evidence and exposing precise paragraph-level or relational evidence trails, the Polymer Literature Scholar framework strengthens trust in AI-generated responses and enables reliable synthesis of knowledge from an increasingly complex polymer literature. While demonstrated here for PHAs, the approach is broadly applicable to other materials domains where scientific insight depends on integrating fragmented and heterogeneous evidence.

\section*{Methods}\label{sec:methods}

\subparagraph{Corpus.}\label{sec:methods_corpus}
Our literature corpus comprises $\sim$ 3 million documents obtained from publishers including Elsevier, Wiley, Springer Nature, the American Chemical Society, and the Royal Society of Chemistry, covering articles published up to 2025. Details of the parsing pipeline are provided in \cite{gupta2024data}.  

To construct the PHA-specific corpus, we applied a keyword-based filter to the titles and abstracts of all articles, using PHA-relevant terms listed in the Supplementary Information. The filtered results were then manually verified, producing a final set of 1,028 PHA-relevant papers. The paragraphs were subsequently extracted from the full texts of these articles for downstream processing.

\subparagraph{Evaluation Metrics.}\label{sec:methods_eval_metrics}
Retrieval performance was evaluated using Recall@K,  Recall PID@K and Accuracy. Recall@K is a standard metric in information retrieval for measuring the accuracy and ranking quality of retrieved results. It quantifies whether the expected ground-truth context paragraph appears among the top-$K$ retrieved results.

In scientific literature, relevant information is often distributed across multiple paragraphs within the same paper. As a result, Recall@K may underestimate retrieval performance when the correct paper is retrieved but the exact ground-truth paragraph is not. To address this limitation, we additionally report Recall PID@K, which considers a retrieval correct if any of the top-$K$ retrieved contexts originate from the ground-truth paper, irrespective of the specific paragraph matched.

Because both Recall@K and Recall PID@K focus solely on retrieval behavior, they may still underestimate the practical usefulness of a RAG system. We therefore also report Accuracy, defined as a binary (0/1) score assigned by a domain expert after assessing the correctness and completeness of the generated answer with respect to the query.

\subparagraph{Recall@K.}
Recall@K measures whether the expected ground-truth paragraph $P_{\text{expected}}$ is included within the top-$K$ retrieved paragraphs $P_{\text{retrieved}}$:
\begin{equation}
\text{Recall@K} =
\begin{cases}
1, & \text{if } P_{\text{expected}} \in P_{\text{retrieved}}[:K], \\
0, & \text{otherwise}.
\end{cases}
\end{equation}
For $N$ total queries, the mean Recall@K is:
\begin{equation}
\text{Mean Recall@K} = \frac{1}{N} \sum_{i=1}^{N} \text{Recall@K}_i.
\end{equation}

\subparagraph{Recall PID@K.}
Recall PID@K measures whether the paper containing the ground-truth paragraph, identified by its Paragraph ID ($\text{PID}_{\text{expected}}$), is included within the top-$K$ retrieved papers ($\text{PID}_{\text{retrieved}}$):
\begin{equation}
\text{Recall PID@K} =
\begin{cases}
1, & \text{if } \text{PID}_{\text{expected}} \in \text{PID}_{\text{retrieved}}[:K], \\
0, & \text{otherwise}.
\end{cases}
\end{equation}
The mean Recall PID@K is computed analogously to Equation~(2).


\subsection*{Knowledge Graph Construction Phase}
\subparagraph*{Entity Extraction.}
Knowledge graph tuples were extracted from the corpus using LLMs via in-context learning. Each extracted tuple contains five fields: \emph{(subject, relation, object, reference\_relation, reference\_node)}. When available in the source text, explicit citations and figure/table references are also recorded. The optimized prompt used for tuple generation is provided in the Supplementary Information.


\subparagraph*{Entity Canonicalization.}
To unify semantically equivalent entities among the extracted tuples, 384-dimensional sentence embeddings were generated using the \texttt{all-MiniLM-L6-v2} model from the SentenceTransformers library. A two-stage hybrid clustering strategy was employed to balance scalability and precision. First, MiniBatchKMeans was used to perform coarse-grained grouping of entities into 2,000 clusters. Within each cluster, AgglomerativeClustering with average linkage was applied to perform fine-grained merging based on a cosine distance threshold of 0.5.

For each resulting cluster, the canonical entity was selected using centroid-based ranking, where the entity closest to the cluster centroid was chosen as the representative label. For clusters dominated by numerical entities (defined as clusters with over 60\% numeric content), a synthetic canonical label of the form \texttt{numerical\_value} was assigned. This canonicalization step reduces vocabulary fragmentation while preserving semantic relationships, thereby improving graph connectivity and retrieval robustness.

\subsection*{GraphRAG Retrieval Phase}

\subparagraph*{Query Preprocessing.}
User queries undergo a preprocessing step to extract domain-relevant terms and normalize textual variations prior to retrieval. Queries are first lowercased and processed using tokenization and lemmatization implemented with the NLTK framework. A custom regular expression pattern is then applied to identify domain-specific entities, including polymer names, chemical formulas, composite material notations, and numerical values. Stop words are removed while preserving scientifically meaningful terms. This preprocessing standardizes query representations and facilitates robust matching against entities in the knowledge graph.

\subparagraph*{String-Based Retrieval.}
String-based retrieval identifies candidate knowledge graph tuples through exact keyword matching. Regular expressions with word-boundary constraints are applied to the subject, relation, and object fields of each tuple, with matching performed in a case-insensitive manner. Tuples containing all query keywords are prioritized. If no such tuples are found, a relaxed matching strategy is applied, requiring at least two keyword matches. Retrieved tuples are ranked using a frequency based scoring scheme that prioritizes tuples matching a larger number of query terms.

\subparagraph*{Canonical-Based Retrieval.}
To account for lexical variation in the literature, semantic retrieval is performed using canonical entity embeddings produced as part of the entity normalization step. Query terms are embedded using the \texttt{all-MiniLM-L6-v2} model and compared against precomputed canonical entity embeddings using cosine similarity. Canonical entities exceeding a similarity threshold of 0.7 are retained, and all tuples containing these entities are retrieved. Exhaustive similarity comparison across the canonical entity set ensures coverage of semantically related tuples that may not be captured through string-based matching alone.

\subparagraph*{Hybrid Retrieval and Re-ranking.}
Hybrid retrieval integrates string-based and canonical-based matching to leverage both exact keyword overlap and semantic similarity. For each tuple, a hybrid score $S_{\text{hybrid}}$ is computed as a weighted combination of the string-match score $S_{\text{string}}$ and the canonical-match score $S_{\text{canonical}}$:
\begin{equation}
S_{\text{hybrid}} = \alpha \cdot S_{\text{canonical}} + (1 - \alpha) \cdot S_{\text{string}},
\end{equation}
where $\alpha = 0.7$, reflecting greater emphasis on semantic similarity. Hybrid scores are min–max normalized, and tuples with $S_{\text{hybrid}}$ below a threshold of $\tau = 0.6$ are filtered out.

\subparagraph*{Path Re-ranking.}
The retained candidate tuples after hybrid matching are re-ranked using a cross-encoder model, \texttt{cross-encoder/ms-marco-MiniLM-L-6-v2}, which assigns a re-ranking score $S_{\text{rerank}}$ by jointly encoding each query–tuple pair. Unlike entity-level matching, this path-level evaluation considers the complete relational path (subject, relation, object) in the context of the query, enabling the model to assess semantic coherence and relevance across the full tuple. This step improves retrieval precision by prioritizing tuples that more directly and collectively address the query.

The final ranking score is computed as:
\begin{equation}
S_{\text{final}} = \lambda \cdot S_{\text{rerank}} + (1 - \lambda) \cdot S_{\text{hybrid}},
\end{equation}
where $\lambda = 0.7$ biases the final ranking toward the context-aware re-ranking score.

The number of tuples returned is controlled by the \texttt{max\_tuples} parameter. To avoid prematurely discarding potentially relevant candidates, the cross-encoder evaluates up to four times \texttt{max\_tuples}, normalizes the resulting scores to the range $[0,1]$, and retains the top \texttt{max\_tuples} tuples. We evaluated \texttt{max\_tuples} values between 300 and 500 and found that 300 provides the best balance between retrieval quality, computational cost, and inference latency.

\subsection*{VectorRAG}

\subparagraph*{Embedding Model.}
VectorRAG employs the \texttt{Qwen/Qwen3-Embedding-4B} model from the SentenceTransformers library to generate 2,560-dimensional embeddings for both the corpus and the user query. Corpus embeddings are pre-computed and stored in a PostgreSQL database for efficient retrieval.

\subparagraph*{Text chunking strategy.}
Scientific articles are typically organized into major sections such as Introduction, Methods, Results and Discussion, and Conclusion, each of which may contain multiple hierarchical subsections. To preserve the scientific context during retrieval, paragraphs were combined based on this structural hierarchy. Specifically, all paragraphs belonging to the same first-level subsection (i.e., the lowest common subheading under a major section) were merged into a single chunk. This approach maintained the logical flow and coherence of scientific arguments within a section while optimizing semantic understanding and retrieval performance.


\section*{Data Availability}
Data sharing is not applicable to this article as no new data was created or analyzed in this study.

\section*{Code Availability}
The code used in this work can be found at \url{https://github.com/Ramprasad-Group/RAG}

\section*{Acknowledgement}
This work was supported by the Office of Naval Research through grants N00014-19-1-2103 and N00014-20-1-2175.

\section*{Competing Interests}
The authors declare no competing interests.

\section*{Supporting Information}
The online version of this article contains Supplementary Information available at \url{https://doi.org/}
\begin{itemize}
    \item Optimized prompts for VectorRAG and GraphRAG components, LLM Benchmarking, Domain Expert Evaluation Questions etc.
\end{itemize}

\def\bibsection{\section*{References}}
\bibliography{citations}
\end{document}